\documentclass[12pt]{article}
\usepackage{graphicx}
\usepackage{amssymb}
\usepackage{amscd}
\usepackage{amsmath}
\usepackage{appendix}

\begin{document}
\begin{titlepage}

{\hbox to\hsize{\hfill revised July 2014 }}

\bigskip \vspace{3\baselineskip}

\begin{center}
{\bf \large 

Inflationary Baryogenesis in a Model with Gauged Baryon Number}

\bigskip

\bigskip

{\bf Neil D. Barrie and Archil Kobakhidze \\ }

\smallskip

{ \small \it
ARC Centre of Excellence for Particle Physics at the Terascale, \\
School of Physics, The University of Sydney, NSW 2006, Australia \\
E-mails: nbar5465@uni.sydney.edu.au, archilk@physics.usyd.edu.au
\\}

\bigskip
 
\bigskip

\bigskip

{\large \bf Abstract}

\end{center}
\noindent 
We argue that inflationary dynamics may support a scenario where significant matter-antimatter asymmetry is generated from initially small-scale quantum fluctuations that are subsequently stretched out over large scales. This scenario can be realised in extensions of the Standard Model with an extra gauge symmetry having mixed anomalies with the electroweak gauge symmetry. Inflationary baryogenesis in a model with gauged baryon number is considered in detail. 
 
\end{titlepage}

\baselineskip=16pt

\section{Introduction}

It is a standard lore that dynamical generation of matter-antimatter asymmetry must happen after an inflationary epoch, since any asymmetry generated before that is diluted away due to the rapid spacetime expansion. In order to produce a significant asymmetry, during inflation, the production rate of baryonic charge must exceed its dilution rate. Actually, inflationary dynamics may support such a scenario: if a large baryonic charge density is created due to small-scale quantum fluctuations, it will typically be stretched out over large scales due to inflation. This basic observation has been realised in a model of inflationary leptogenesis \cite{Alexander:2004us}, where a lepton asymmetry is produced during inflation due to the gravitational birefringence through a gravitational lepton number anomaly coupled to an extra pseuodoscalar field. 

In this paper we argue that the inflationary baryogenesis scenario can be realised in extensions of the Standard Model with an anomalous gauge symmetry which have mixed anomalies with electroweak gauge symmetry.\footnote{In the early universe, when the expansion rate is faster than processes with fermion chirality flip, the gauged anomaly may effectively appear within the Standard Model \cite{Campbell:1992jd}. Indeed, it has been argued in \cite{Giovannini:1997eg}  that anomalous production  of the right-handed electron number is possible through the hypercharge anomaly. An inflationary version of the above scenario is discussed in \cite{Alexander:2011hz}.} This anomalous theory can also be viewed as an effective low-energy theory, which admits a fundamental completion free of gauge anomalies. The obvious candidates for such an anomalous gauge theory are gauged baryon ($B$) and lepton ($L$) numbers, or any linear combination thereof except for $(B-L)$. In the present paper we consider a model with gauged $B-$number in detail.

The basic three Sakharov's conditions for dynamical baryogenesis \cite{Sakharov:1967dj} are satisfied in our model as follows. As in the Standard Model the baryon number is not conserved because of the mixed electroweak - $B$ anomaly. On top of this, $U(1)_B$ gauge invariance requires a pseudoscalar field, that describes the longitudinal polarization of the baryonic photon, to couple to the anomaly. In the cosmological setting these interactions spontaneously violate ${\cal CP}$ invariance and lead to the ${\cal  CP}-$asymmetric out-of-equilibrium production of  electroweak gauge bosons with different polarizations. In particular, during an inflationary epoch the produced particles form a Bose-Einstein condensate with a large correlation length which supports the generation of a non-zero baryon number through the anomaly.

The rest of the paper is organized as follows. In the next section we describe a model with gauged $B-$number. In sec. 3, we present quantization of the weak gauge bosons in an inflationary spacetime. In sec. 4, we compute the generated baryon asymmetry. Sec. 5 is reserved for conclusions. Finally, some technical details of our calculations and useful formulas are delegated to appendices \ref{a} and \ref{b}.            
     
 \section{A model with gauged $B-$number}

Let us consider an extension of the Standard Model with gauged symmetry $SU(3)\times SU(2)\times U(1)_Y\times U(1)_B$. We assume that no extra fermions and scalars are introduced beyond those in the Standard Model. Since the gauged baryon number $U(1)_B$ is anomalous, the associated gauge boson carries three degrees of freedom \cite{Preskill}, that is, it is necessarily massive. A scalar field $\theta (x)$ that describes the longitudinal degree of freedom of a massive baryonic photon $X_{\mu}$ can be used to cancel out anomalies  without introducing new matter fermions \cite{Faddeev:1986pc, Preskill}. The addition to the Lagrangian density, describing the Standard Model, then reads:
\begin{eqnarray}
\frac{1}{\sqrt{-g}}{\cal L}_B=-\frac{1}{4}g^{\mu\alpha}g^{\nu\beta}X_{\mu\nu}X_{\alpha\beta}+\frac{1}{2}f_B^2g^{\mu\nu}\left(g_BX_{\mu}-\partial_{\mu}\theta \right)\left(g_B X_{\nu}-\partial_{\nu}\theta \right) \nonumber \\
+\frac{3\theta(x)}{32\pi^2}\left[g_1^2B_{\mu\nu}\tilde B^{\mu\nu}-g_2^2W^{a}_{\mu\nu}\tilde W^{a\mu\nu}\right]  
\label{1}
\end{eqnarray}    
where $X_{\mu\nu}$, $B_{\mu\nu}$ and $W^{a}_{\mu\nu}$ ($a=1,2,3$; summation under the repeated weak isospin indices is assumed throughout the paper) denote field strengths for $U(1)_B$, $U(1)_Y$ and $SU(2)$ gauge bosons with corresponding coupling constants  $g_B$,~$g_1$, and $g_2$, respectively; $f_B$ is a parameter that defines the mass of the baryonic photon, $m_B=g_Bf_B$; $\tilde B^{\mu\nu} (\tilde W^{a\mu\nu}) =\frac{1}{2\sqrt{-g}}\epsilon^{\mu\nu\rho\sigma}B_{\rho\sigma }(W^{a}_{\rho\sigma})$ is the dual field strength, and $\epsilon^{\mu\nu\rho\sigma}$ is the Levi-Civita tensor.  

Note, that the terms in the second line of Eq. (\ref{1}) are introduced to maintain gauge invariance of the full quantum theory under $U(1)_B$ transformations. Indeed, while they are not invariant under $U(1)_B$ gauge transformations, $X_{\mu}\to X_{\mu}+(1/g_B)\partial_{\mu}\alpha$ and $\theta(x)\to \theta(x)+\alpha(x)$, their variance cancels out against the gauge variation of the functional measure of quark fields within the path integral quantization framework. It is clear that the above model can also be viewed as an effective low-energy approximation of an anomaly-free theory \cite{Foot:1989ts}, where additional fermionic fields, 
which cancel the $[SU(2)]^2-U(1)_B$ and $[U(1)_Y]^2-U(1)_B$ mixed anomalies, are integrated out. Then, according to t'Hooft's anomaly matching condition \cite{tHooft}, the terms restoring gauge invariance necessarily appear in the low-energy theory.

A remark related to the above $U(1)_B$ gauge invariance is in order. In principle one may locally fix the gauge such that $\theta(x)=0$,\footnote{There may exist a topological obstruction to imposing this gauge condition globally in spacetime because of the presence of vortex excitations around which $\theta(x)$ has a nontrivial winding number. However, within the perturbative framework this complication is irrelevant; hence we ignore this non-perturbative effect here.} so that the theory with $\theta$ field is equivalent (within the perturbation theory) to a theory with purely massive $X_{\mu}$ coupled to quarks without the $\theta$ field (`unitary gauge'). Nevertheless, we find it to be more convenient if  $\theta$ is manifestly present as in Eq. (\ref{1}), since the longitudinal physical degree of freedom of the massive baryonic photon, which plays a crucial role in our analysis, is easily identifiable in this case.

The metric tensor in Eq. (\ref{1}) describes a homogeneous and spatially flat cosmological spacetime, and hence, in conformal coordinates can be written as: $g_{\mu\nu}=a^2(\tau)\eta_{\mu\nu}$ and $g\equiv {\rm det (g_{\mu\nu})}$. The scale factor $a(\tau)$ during inflation reads:
\begin{equation}
a(\tau)=-1/H_{\rm inf}\tau~,
\label{2}
\end{equation} 
where $H_{\rm inf}$ is an expansion rate ($H_{\rm inf}\cong {\rm const.}$) and $\tau \in [-\infty, 0]$ is the conformal time.  
 
 To proceed further we make the following simplifying assumptions.  We assume that $g_B\ll 1$, and thus $\theta (x)$ and $X_{\mu}$ fields essentially decouple from each other. The smallness of the $U(1)_B$ coupling constant implies the baryonic photon is relatively light, $m_B/f_B\ll 1$, and hence we will not be interested in its dynamics during inflation. We also ignore the dynamics of the hypercharge gauge field $B_{\mu}$ as it is less relevant compared to the dynamics of weak isospin fields $W_{\mu}^{a}$, due to the fact that $g_2 > g_1$. Furthermore, as we are interested in small quantum fluctuations of $SU(2)$ gauge bosons around a trivial (vacuum) configuration, we ignore self-interactions of $W_{\mu}^{a}$ restricting to the linearized approximation. For the $\theta(x)$ field we only consider a classical homogeneous background configuration, $\theta(\tau, \vec{x})=\theta(\tau)$, and ignore quantum fluctuations over it. With these assumptions the Lagrangian terms being considered significantly simplify to:
\begin{equation}
{\cal L}=-\frac{1}{4}\eta^{\mu\rho}\eta^{\nu\sigma}W^{a}_{\mu\nu}W^{a}_{\rho\sigma}+\frac{a^2(\tau)}{2}(\phi'(\tau))^2-
\frac{3g_2^2}{64\pi^2f_B}\phi(\tau)\epsilon^{\mu\nu\rho\sigma}W^{a}_{\mu\nu}W^{a}_{\rho\sigma}~,  
\label{3}
\end{equation} 
where $\phi(\tau)\equiv f_B\theta(\tau)$ and $\phi'\equiv d\phi/d\tau$.                

The equation of motion for $\phi(\tau)$ that follows from the above Lagrangian reads:
\begin{equation}
\left(a^2\phi'\right)'=0~,
\label{4}
\end{equation}
where we have ignored terms quadratic in $W_{\mu}^{a}$. From Eq. (\ref{4}) we obtain:
\begin{equation}
\phi'(\tau)=\frac{\phi'_0}{a^2(\tau)}~,
\label{5}
\end{equation}
where $\phi'_0$ is an integration constant associated with the `field velocity' at the start of inflation $\tau=\tau_0,~a(\tau_0)=1$. Plugging Eq. (\ref{5}) into the linearized equation of motion for the $W_{\mu}^{a}$ gauge fields we obtain:
\begin{equation}
\left(\partial_{\tau}^2-\vec \bigtriangledown^2\right)W^{ai}+\kappa \tau^2 \epsilon^{ijk}\partial_jW^{a}_k=0~,
\label{6}
\end{equation}
where 
\begin{equation}
\kappa =\frac{3g_2^2\phi'_0H_{\rm inf}^2}{8\pi^2f_B}~,
\label{6a}
\end{equation} 
and we have adopted the gauge 
where $W_0^{a}=\partial_iW_i^{a}=0$.  Note that the first and the last terms in Eq. (\ref{6}) have opposite ${\cal P}$ and, hence, ${\cal CP}$ parities. This is the source of ${\cal CP}$ violation in our model which is one of the necessary Sakharov's conditions \cite{Sakharov:1967dj}.  

\section{Quantum fluctuations of the weak gauge bosons during inflation}
To quantize the model described in the previous section we promote the weak gauge boson fields to operators:
\begin{equation}
W^{a}_i=\int\frac{d^3\vec k}{(2\pi)^{3/2}}\sum_{\alpha}\left[F_{\alpha}(\tau,k)\epsilon_{i\alpha}\hat a_{\alpha}^{a}{\rm e}^{i\vec k\cdot\vec x}+
F^{*}_{\alpha}(\tau,k)\epsilon^{*}_{i\alpha}\hat a_{\alpha}^{a\dagger}{\rm e}^{-i\vec k\cdot\vec x}
\right]~,
\label{7}
\end{equation}
where creation, $\hat a_{\alpha}^{a\dagger}(\vec k)$, and annihilation, $\hat a_{\alpha}^{a}(\vec k)$, operators satisfy canonical commutation relations:
\begin{equation}
\left[\hat a_{\alpha}^{a}(\vec k), \hat a_{\beta}^{b\dagger}(\vec k')\right]=\delta_{\alpha\beta}\delta^{ab}\delta^3(\vec k-\vec k')~,
\label{8}
\end{equation}
 and      
\begin{equation}
\hat a^a_{\alpha}(\vec k)\vert 0\rangle_{\tau}=0~,
\label{14}
\end{equation}
where $ \vert 0\rangle_{\tau} $ is an instantaneous vacuum state at time $ \tau $.

In Eq. (\ref{7}), two vectors $\vec \epsilon_{\alpha}$ ($\alpha=+,-$) describe two helicity states (we treat the weak bosons as massless particles, since $m_W<<H_{\rm inf}$) and they are in fact complex conjugates of each other, i.e. $\vec \epsilon_{+}^{*}=\vec \epsilon_{-}$.  The equations for the mode functions, $F_{\pm}(\tau,k)$ [$k\equiv|\vec k|$], straightforwardly follow from Eq. (\ref{6}):
\begin{equation}
F''_{\pm}+\left(k^2\mp \kappa\tau^2 k \right)F_{\pm}=0~. 
\label{9}
\end{equation}  
According to this equation, towards the end of inflation ($ \tau_{\rm end} \simeq 0 $) all the modes with $k>>\mu=|\kappa| \tau_{\rm end}^2$ approach   
${\cal CP}-$symmetric flat spacetime plane waves:
\begin{equation}
F_{\pm}(\tau,k)\stackrel{\tau \rightarrow 0}{\longrightarrow} \frac{1}{\sqrt{2k}}~. 
\label{10}
\end{equation}  
These also include large wavelength superhorizon modes $ k\vert \tau_{\rm end} \vert \ll 1$, which are of our prime interest. The field operator Eq. ($\ref{7} $) for $ \tau \rightarrow 0 $ becomes:

\begin{equation}
W^{a}_i=\int\frac{d^3\vec k}{(2\pi)^{3/2}\sqrt{2k}}\sum_{\alpha}\left[\epsilon_{i\alpha}\hat b_{\alpha}^{a}{\rm e}^{-ik|\tau| + i\vec k\cdot\vec x}+
\epsilon^{*}_{i\alpha}\hat b_{\alpha}^{a\dagger}{\rm e}^{ik|\tau| -i\vec k\cdot\vec x}
\right]~.
\label{13}
\end{equation}

 The nonzero term $\propto \kappa\tau^2 k$ in Eq. (\ref{9}) is responsible for ${\cal CP}-$asymmetric ($F_+\neq F_-$) solutions:
\begin{equation}
F_{+}(\tau,k)=C_1 D_{-\frac{1}{2}\left(1-\Omega_k\right)}\left(\frac{\sqrt{2}k\tau}{\sqrt{\Omega_k}}\right)+
C_2 D_{-\frac{1}{2}\left(1+\Omega_k\right)}\left(\frac{i\sqrt{2}k\tau}{\sqrt{\Omega_k}}\right)~,
\label{11}
\end{equation} 
and 
\begin{equation}
F_{-}(\tau,k)=C_3 D_{-\frac{1}{2}\left(1+i\Omega_k\right)}\left(\frac{\sqrt{2i}k\tau}{\sqrt{\Omega_k}}\right)+
C_4 D_{-\frac{1}{2}\left(1-i\Omega_k\right)}\left(\frac{i\sqrt{2i}k\tau}{\sqrt{\Omega_k}}\right)~,
\label{12}
\end{equation}
where $D_{\nu}(z)$ is the parabolic cylinder function and $\Omega_k=\left(\frac{k^3}{\kappa}\right)^{1/2}$. The integration constants $C_{1,2,3,4}$ are defined through the Wronskian normalization condition and by matching Eqs. (\ref{11},\ref{12}) with plane wave modes according to Eq. (\ref{10}). For superhorizon modes ($k|\tau|\to 0$), which are of our prime interest, they are given in Appendix \ref{a}, Eqs. (\ref{a1}-\ref{a4}).

Two sets of creation and annihilation operators, $\lbrace \hat a_{\alpha}^a, \hat a^{a\dagger}_{\alpha} \rbrace$ and $\lbrace\hat b^a_{\alpha}, \hat b^{a\dagger}_{\alpha} \rbrace$, in Eqs ($ \ref{7} $) and ($ \ref{13} $), are related through the Bogoliubov transformations:
\begin{eqnarray}
\label{15}
\hat b^a_{\alpha}(\vec k)= \alpha_{\alpha} a^{a\dagger}_{\alpha}(\vec k)+\beta^{*}_{\alpha}\hat a_{\alpha}^{a}(\vec k) \\
\label{16}
\hat b^{a\dagger}_{\alpha}(\vec k)= \alpha^{*}_{\alpha} a^{a}_{\alpha}(\vec k)+\beta_{\alpha}\hat a_{\alpha}^{a\dagger}(\vec k)
\end{eqnarray}
The Bogoliubov coefficients for the superhorizon modes ($k|\tau_{\rm end}|\approx 0$) of interest can be computed explicitly:
\begin{equation}
\alpha_{\alpha}=\frac{1}{2}+i\sqrt{\frac{1}{2k}}R^*_{\alpha}~~{\rm and}~~
\beta_{\alpha}=\frac{1}{2}-i\sqrt{\frac{1}{2k}}R^*_{\alpha}~.
\label{17}
\end{equation}
where $ R^*_{\alpha}:=F^{*\prime}_{\alpha}\vert_{\frac{\kappa\tau_{\rm end}^2}{k},k|\tau_{\rm end}|\to 0} $.
   
\section{Computing the baryon asymmetry}
We are now ready to compute the generated baryon number density. Anomalous non-conservation of baryonic current
\begin{equation}
\partial_{\mu}\left(\sqrt{-g} j_B^{\mu}\right)=\frac{3g_2^2}{64\pi^2}\epsilon^{\mu\nu\rho\sigma} 
W^{a}_{\mu\nu} W^{a}_{\rho\sigma}\equiv \frac{3g_2^2}{16\pi^2}\partial_{\mu}\left(\sqrt{-g} K^{\mu}\right)~,
\label{18}
\end{equation}
where $K^{\mu}=\frac{1}{2\sqrt{-g}}\epsilon^{\mu\nu\rho\sigma}W^{a}_{\nu\rho}W^{a}_{\sigma}$ is a topological current, implies that the net baryon number density $n_B=n_b-n_{\bar b}\equiv a^{-1}(\tau)\langle 0 \vert j_B^0\vert 0 \rangle$ is related to the weak gauge boson Chern-Simons number density, $n_{CS}=_{\tau_0}\langle 0\vert K^{0}(\tau)\vert 0\rangle_{\tau_0}$,  at the end of inflation, $\tau=\tau_{\rm end}$:
\begin{eqnarray}
n_B= \frac{3g_2^2}{16\pi^2} a(\tau_{\rm end}) n_{CS}~.
\label{19}
\end{eqnarray}
Here, $n_B(\tau_0)= n_{CS}(\tau_0)=0$, at the start of inflation. Furthermore, we are interested in $n_{CS}$ for large scale superhorizon modes ($k|\tau|\approx 0$), hence, we have:
\begin{eqnarray}
n_{CS}= \frac{1}{a^4(\tau_{\rm end})}\epsilon^{ijk}\lim_{k|\tau|\to 0}\langle 0\vert W_i \partial_j W_k\vert 0\rangle 
=\frac{3}{8\pi^2 a^4(\tau_{\rm end})}\int_\mu^{\Lambda}k dk\left[\left\vert R_{+}\right \vert^2-\left \vert R_{-}\right \vert^2\right]~, 
\label{20}
\end{eqnarray}
where 
\begin{eqnarray}
\label{21}
\left\vert R_{+}\right \vert^2=\frac{\pi}{2}\sqrt{\frac{\kappa k}{2}}\left\vert C_1\frac{2^{\frac{\Omega_k}{4}}(1-\Omega_k)}{ \Gamma\left(\frac{5-\Omega_k}{4}\right)}+
iC_2\frac{(1+\Omega_k)}{ 2^{\frac{\Omega_k}{4}}\Gamma\left(\frac{5+\Omega_k}{4}\right) }
\right|^2~, \\ \nonumber \\
\left\vert R_{-}\right \vert^2=\frac{\pi}{2}\sqrt{\frac{\kappa k}{2}}\left\vert
C_3\frac{(1+i\Omega_k)}{ 2^{\frac{i\Omega_k}{4}}\Gamma\left(\frac{5+i\Omega_k}{4}\right)}+
iC_4\frac{2^{\frac{i\Omega_k}{4}}(1-i\Omega_k)}{\Gamma\left(\frac{5-i\Omega_k}{4}\right)}
\right\vert^2~,
\label{22}
\end{eqnarray}
and $\Lambda$ is an ultraviolet cut-off, and $ \mu $ is an IR cut-off. We have found that the integral in Eq. (\ref{20}) is dominated by the dependence on $ \mu $ given below, and is independent of $\Lambda$.  This result can be understood as follows. Physically, the modes with large $k$ are essentially ${\cal CP}-$invariant plane waves, thus the integrand in Eq. (\ref{20}) for those modes nullifies. Thus, the integral is effectively zero for large $ k $ modes. The IR cut-off is naturally given by $ \mu=\kappa \tau_{\rm end}^2 $ which corresponds to the modes that were initially matched to the Minkowski planewave solutions, in Eq. (\ref{10}).

Finally, assuming that there was no significant entropy production after the reheating phase, we estimate the entropy density as: $s \simeq \frac{2\pi^2}{45}g^{*}T^3_{\rm rh}$, where $ g^{*}(T_{\rm rh}) \sim 100$ and $T_{\rm rh}$ is the reheating temperature. We obtain the following simple expression for the baryon asymmetry parameter:

\begin{equation}
\eta_B=\frac{n_B}{s}\simeq  \frac{5 g_{2}^{2}}{g^{*}\sqrt{2}\pi^{7}} \frac{\Gamma\left(\frac{3}{4}\right)^4}{\Gamma\left(\frac{5}{4}\right)^2}e^{-3 N_{\rm e}}\left(\frac{\kappa}{ \mu T_{\rm rh}^2}\right)^\frac{3}{2}\simeq 4.1\cdot 10^{-3} \frac{H_{\rm inf} T_{\rm rh}}{M_{p}^{2}}~,
\label{24}
\end{equation}
where $ \tau_{\rm end} = -\frac{1}{a(\tau_{\rm end}) H }= -\frac{e^{-N_{\rm inf}}}{H_{\rm inf} } $ and $g^2_2\approx 4\pi/29$. The total number of e-folds $N_e$, that defines the dilution factor, includes the minimal number of e-folds required during inflation $N_{\rm inf}\simeq 34+\ln\left(\frac{T_{\rm rh}}{100~{\rm GeV}}\right)$ and the number of e-folds during reheating $N_{\rm rh}\simeq \frac{1}{3}\ln\left(\frac{45H^2_{\rm inf}M^2_{p}}{4\pi^3 g^*T_{\rm rh}^4}\right)$:
\begin{eqnarray}
N_e= N_{\rm inf} + N_{\rm rh} \simeq 32 + \ln\left(\frac{T_{\rm rh}}{100~{\rm GeV}}\right)+\frac{2}{3}\ln\left(\frac{H_{\rm inf}M_{p}}{T_{\rm rh}^2}\right)
\label{25}
\end{eqnarray}

 Eq. (\ref{24}) was obtained using a first order Taylor expansion around $ \Omega_{k}=0 $. Interestingly, for the chosen IR cut-off $ \mu=|\kappa|\tau_{end}^2 $, the asymmetry parameter is not manifestly dependent on $ \kappa $, due to the approximation adopted in our calculations. Indeed, in the opposite limit of vanishing $ \kappa \rightarrow 0 $ and $ \Omega_k \rightarrow \infty $ leads to the $ F_{\pm} $ solutions to approach the flat spacetime limit, where the resulting asymmetry is 0. From Eq (\ref{24}), the following requirement is obtained:
 \begin{equation}
 H_{\rm inf} T_{\rm rh} \simeq 3 \times 10^{30} ~\textrm{GeV}^2~.
 \label{27}
 \end{equation}
 Hence, the desired value of $\eta_B\approx 8.5\cdot 10^{-11}$ can be obtained as long as the Hubble rate and reheating temperature are suitable large as to satisfy Eq. (\ref{27}) (i.e. $ H\sim 10^{14} $ GeV and $ T_{\rm rh} \sim 10^{16} $ GeV).

The net baryon number density $n_B$ Eq. (\ref{19}) generated during inflation evolves in the subsequent epochs. Besides the trivial dilution due to the expansion, which is cancelled out in the asymmetry parameter Eq. (\ref{24}), there may be other processes that influence $n_B$. For example, non-perturbative $(B+L)$-violating processes, which are thermally activated if $T_{\rm rh}\gtrsim 100$ GeV \cite{Kuzmin:1985mm},   wash out any existing $(B+L)$ number, while preserving $(B-L)$ in thermal equilibrium. This means that part of the initial baryon number will be reprocessed into a lepton number, but $n_B$ will remain of the same order of magnitude.

\section{Conclusion} 

In this paper we have argued that a successful baryogenesis scenario can be realised during the inflationary epoch within a class of anomalous gauge theories. A model with gauged baryon number has been considered in detail. The large wavelength modes of electroweak gauge bosons, produced during inflation, form a Bose-Einstein condensate that supports non-zero net baryon number density $n_B$. We have found that the baryon number asymmetry parameter $\eta_B$ has a simple dependence Eq. (\ref{24}) on the cosmological parameters $H_{\rm inf}$ and $ T_{\rm rh}$ Eq. (\ref{27}), for which the experimental values can be accommodated. To obtain the desired asymmetry large scale inflation $ H\sim 10^{14} $ GeV and high reheating temperature $ T_{\rm rh} \sim 10^{16} $ GeV are required. This is in accord with indications on the inflationary scale from the BICEP2 measurements of B-modes \cite{Ade:2014xna}.  

Several different versions of the model presented here are also possible. In fact, any model with an additional gauge symmetry having mixed anomalies with the electroweak symmetry can potentially provide a successful framework for inflationary baryogenesis. An interesting aspect of these class of models is that hypothetical new physics behind the baryogenesis scenario may well be accessible at the LHC. It will be interesting to study collider phenomenology of these models as well.

\paragraph{Acknowledgment.} This work was partially supported by the Australian Research Council. 

\appendix
\section{Appendix}
\label{a}

\protect{\subsubsection*{$ F_{+} $ Coefficients, Eq. (\ref{11})}}

Matching superhorizon modes with the plane waves we obtain the following relation:
\begin{equation*}
 C_{1} =\frac{\Gamma(\frac{3-\Omega_{k}}{4})}{2^{\frac{-1}{4}(1-\Omega_{k})}\sqrt{\pi}}\left( \frac{1} {\sqrt{2k}} - C_{2} \frac{2^{\frac{-1}{4}(1+\Omega_{k})}\sqrt{\pi}}{\Gamma(\frac{3+\Omega_{k}}{4})}\right)
\end{equation*}
The Wronskian normalisation implies:
\begin{equation*}
\sqrt{\frac{2}{\Omega_{k}}}C_{1}C_{2}  \sin(\frac{\pi}{4} (1+\Omega_{k})) + C_{2}^{2}\sqrt{\frac{\pi}{\Omega_{k}}}\frac{1}{\Gamma(\frac{1+\Omega_{k}}{2})}=\frac{1}{2k}
\end{equation*}
Solving the above conditions we find that the coefficients for $ F_{+} $ modes are:

\begin{equation} 
C_{1}=\frac{2^{-\frac{1}{4}(1+\Omega_{k})}\Gamma(\frac{3-\Omega_{k}}{4})}{\sqrt{\pi k}}-\frac{2^{-\frac{1}{2}(\Omega_{k}+3)}\Gamma\left(\frac{1+\Omega_{k}}{4}\right)\Gamma(\frac{3-\Omega_{k}}{4})}{ \Gamma(\frac{3+\Omega_{k}}{4})}\sqrt{\frac{\Omega_{k}}{ \pi k}}
 \label{a1}
\end{equation}

and

\begin{equation} 
C_{2}=
\frac{\Gamma\left(\frac{1+\Omega_{k}}{4}\right)}{2\sqrt{2\pi}} \sqrt{\frac{\Omega_{k}}{ k}}=\frac{\Gamma\left(\frac{1+\Omega_{k}}{4}\right)}{2\sqrt{2\pi }}\left(\frac{k}{\kappa}\right)^{\frac{1}{4}}
 \label{a2}
\end{equation}\\

\protect{\subsubsection*{$ F_{-} $ Coefficients, Eq. (\ref{12})}}

Similarly as above we obtain the following relations from the matching,
\begin{equation*}
 C_{4} =\frac{\Gamma(\frac{3-i\Omega_{k}}{4})}{2^{\frac{-1}{4}(1-i\Omega_{k})}\sqrt{\pi}}\left( \frac{1} {\sqrt{2k}} - C_{3} \frac{2^{\frac{-1}{4}(1+i\Omega_{k})}\sqrt{\pi}}{\Gamma(\frac{3+i\Omega_{k}}{4})}\right)~,
\end{equation*}

and the Wronskian normalisation:
\begin{equation*}
C_{3}^{2} +|C_{4}|^{2} + 2C_{3} e^{\frac{-\pi \Omega_{k}}{4}}\sqrt{2\pi} \mathrm{Im}\left(\frac{\sqrt{i}  C_{4}^{*}}{\Gamma(\frac{1+i\Omega_{k}}{2})}\right)=\frac{e^{\frac{-\pi \Omega_{k}}{4}}}{k}\sqrt{\frac{\Omega_{k}}{2}}
\end{equation*}

These two equation determine the coefficients for $ F_{-} $ modes:
\begin{equation}
C_{3}=\frac{1}{2 \sqrt{2k}P(k)}\left(\sqrt{\Omega_{k}}e^{-\frac{\pi \Omega_{k}}{4}}-\frac{1}{\pi}\left|\Gamma\left(\frac{3-i\Omega_{k}}{4}\right)\right|^2\right)
 \label{a3}
\end{equation}

\begin{equation}
 C_{4} =\frac{\Gamma(\frac{3-i\Omega_{k}}{4})}{2^{\frac{-1}{4}(1-i\Omega_{k})}\sqrt{2\pi k}}\left( 1 - \frac{\sqrt{\pi}}{2^{\frac{1}{4}(5+i\Omega_{k})} P(k)\Gamma(\frac{3+i\Omega_{k}}{4})}\left(\sqrt{\Omega_{k}}e^{-\frac{\pi \Omega_{k}}{4}}-\frac{1}{\pi}\left|\Gamma\left(\frac{3-i\Omega_{k}}{4}\right)\right|^2\right) \right)~,
  \label{a4}
\end{equation}
where

%

\begin{equation*}
P(k)=\frac{2^{3/4}}{\sqrt{\pi}}\left(2\pi e^{-\frac{\pi \Omega_{k}}{4}} \mathrm{Im}\left[\frac{\sqrt{i}}{2^{\frac{i\Omega_{k}}{4}}\Gamma(\frac{1+i\Omega_{k}}{4})} \right]-\mathrm{Re}\left[ \frac{ \Gamma\left(\frac{3-i\Omega_{k}}{4}\right)}{2^{\frac{i\Omega_{k}}{4}}}\right]\right)
\end{equation*}
\\

\section{Appendix}
\label{b} 
Here we collect useful formulas and properties of special functions \cite{NIST-DLMF} used in the main text. The parabolic cylinder function is denoted $ D_{\nu}(z) $. It is related to the confluent hypergeometric cylinder $U$ and Whittaker $W$ functions by the following,
\begin{align*}
D_{\nu}(z)&= 2^{\nu/2 +1/4}z^{-1/2} W_{\nu/2 +1/4,-1/4}\left(\frac{1}{2}z^2\right)\\
&=\frac{2^{\nu/2}(-iz)^{1/4}(iz)^{1/4}}{\sqrt{z}} U\left(-\frac{1}{2}\nu,\frac{1}{2},\frac{1}{2}z^2\right)
\end{align*}
The following relation has been utilised:
$ D_{\nu}(z)=U(-\frac{1}{2}-\nu,z) $

The Wronskian identities for the parabolic cylinder function used are:
\begin{equation*}
\mathcal{W}[ U(a,z),U(a,-z)]=\frac{\sqrt{2\pi}}{\Gamma(\frac{1}{2}+a)}
\end{equation*}

\begin{equation*}
\mathcal{W}[ U(a,z),U(-a,\pm i z)]=\mp i e^{\pm i \pi(\frac{a}{2}+\frac{1}{4})}
\end{equation*}

The derivative of the parabolic cylinder function, in the $ U(a,z) $ formalism, with respect to a variable $ \tau $ is:
\begin{equation*}
\frac{dU(a,z(\tau))}{d\tau}=-\frac{dz}{d\tau}\left[ (a+\frac{1}{2})U(a+1,z) +\frac{z}{2}U(a,z)  \right]
\end{equation*}
  When the argument $z$ is set to zero, the above equation reads:
\begin{equation*}
\frac{dU(a,0)}{d\tau}=\frac{dz}{d\tau} \frac{\sqrt{\pi}}{2^{\frac{1}{2}(a-\frac{1}{2})}\Gamma(\frac{1}{2}(\frac{1}{2}+a))}
\end{equation*}


\end{document}